\documentclass[preprint]{aastex631}

\shorttitle{Parallelization of SyMBA}
\shortauthors{Lau \& Lee}

\begin{document}
\title{Parallelization of the Symplectic Massive Body Algorithm (SyMBA) {\boldmath$N$}-body Code}

\correspondingauthor{Man Hoi Lee}
\email{mhlee@hku.hk}

\collaboration{2}{}
\author[0000-0001-7433-1177]{Tommy Chi Ho Lau}
\affiliation{University Observatory, Faculty of Physics, Ludwig-Maximilians-Universität München, Scheinerstr. 1, 81679 Munich, Germany}
\affiliation{Department of Earth Sciences, The University of Hong Kong\\ Pokfulam Road, Hong Kong}

\author[0000-0003-1930-5683]{Man Hoi Lee}
\affiliation{Department of Earth Sciences, The University of Hong Kong\\ Pokfulam Road, Hong Kong}
\affiliation{Department of Physics, The University of Hong Kong\\ Pokfulam Road, Hong Kong}

\begin{abstract}
Direct $N$-body simulations of a large number of particles, especially in the study of planetesimal dynamics and planet formation, have been computationally challenging even with modern machines. This work presents the combination of fully parallelized $N^2/2$ interactions and the incorporation of the GENGA code's close encounter pair grouping strategy to enable MIMD parallelization of the Symplectic Massive Body Algorithm (SyMBA) with OpenMP on multi-core CPUs in shared-memory environment. SyMBAp (SyMBA parallelized) preserves the symplectic nature of SyMBA and shows good scalability, with a speedup of 30.8 times with 56 cores in a simulation with 5,000 fully interactive particles.
\end{abstract}

\section{Introduction} \label{sec:intro}
\cite{Duncan1998} presented a powerful symplectic $N$-body algorithm, named Symplectic Massive Body Algorithm (SyMBA). It combines a variant of the \cite{Wisdom1991} method with a multiple time step method to handle close encounters. Alternatively, \cite{Chambers1999} presented the hybrid integrator Mercury, which uses a conventional non-symplectic integrator, such as Bulirsch-Stoer, for close encounters (see also Mercurius by \citealt{Rein2019}). Nonetheless, integrating a large number of particles is impractical with serial computing even on modern computers.

This work was motivated by a study of planetesimal disk and planet formation using SyMBA. Parallelization is achieved by parallelizing the $N^2/2$ interactions and adopting the GENGA code's close encounter pair grouping strategy \citep{Grimm2014}, which is implemented in Fortran by invoking OpenMP in shared-memory environment. It is based on the source code of SyMBA retrieved from the SWIFT package's homepage\footnote{\url{https://www.boulder.swri.edu/~hal/swift.html}} in Jan 2019. As only the sequence of operations within the relevant loop is changed, the symplectic nature of SyMBA is preserved. Users can decide the number of cores to be used, which is beneficial in scanning through parameter space with many simulations and a fixed amount of computing resources.

\section{Parallelization of S\lowercase{y}MBA} \label{sec:intro-symba}
In direct $N$-body codes, computing the gravitational interactions between the bodies dominates processor time. One way to shorten such calculation, as implemented in SyMBA, is by assigning a class of field particles, in contrast to massive particles. The latter have full mutual interactions with all particles, while the former only interacts with massive particles. This can be summarized by a nested loop which takes the form of a right-angled trapezoid:
\begin{verbatim}
do i = 1, N_m
    do j = i+1, N
        !particles i and j interact
    enddo
enddo
\end{verbatim}
where {\ttfamily i} and {\ttfamily j} are dummy variables. {\ttfamily N} equals the total number of particles $N$, while {\ttfamily N\_m} equals the number of massive particles $N_m$, if there is at least one field particle in the system, or {\ttfamily N\_m} equals $(N-1)$, if all particles are massive. By splitting the trapezoid into a rectangle and a triangle and mapping the triangle into a rectangle, all iterations in the loop can be conducted in parallel either by OpenMP's reduction operation\footnote{See \url{https://www.openmp.org/spec-html/5.0/openmpsu107.html}} or critical section assignment\footnote{See \url{https://www.openmp.org/spec-html/5.0/openmpsu89.html}}. 

Close encounter pairs are integrated with adaptive time step in SyMBA. which can be time consuming due to frequent encounters. \cite{Grimm2014} illustrated a solution to handle them in parallel by grouping. A slightly different strategy is adopted here, where grouping is incorporated into the close encounter checking process. Encounter pair is put into an appropriate group when found. Threads that have yet to come across encounters can continue to check for close encounters, while a thread that has found a close encounter pair groups it immediately. The encounter groups can be integrated in parallel while each thread integrates the encounter pairs within the same group sequentially.

``Embarrassing parallelization'' is also applied to simple loops in SyMBA where appropriate. The resulting parallelized version of SyMBA is hereby named SyMBAp.

\section{Benchmarking \& Tests} \label{sec:bm_test}

Following \cite{Abod2019}, a planetesimal disk is generated with the mass $m$ of each particle randomly drawn from a cumulative distribution $N_{>m}=Cm^{-0.3}\exp(-m/m_0)$ with $m_0=10^{-2}M_\oplus$ and minimum mass $m_{\min}=10^{-3} M_\oplus$. The bulk density is assumed to be $2\text{ g cm}^{-3}$. Four different numbers of particles $N$ are chosen: 1,000 (``1k''), 2,000 (``2k''), 5,000 (``5k''), and 10,000 (``10k''). The semimajor axis is randomly drawn from $5-20$ AU. The eccentricity and inclination are randomly drawn from Rayleigh distributions with the scale parameters equal to $10^{-6}$ and $5\times10^{-7}$ respectively. All simulations are integrated with full interaction for $10^4$ years with a time step of $0.5$ year. The time step subdivision factor $M=4$ and the ratio of the radii of adjacent time step subdivision shells $R_k/R_{k+1}=2.08$ are adopted as in the original SyMBA (see \citealt{Duncan1998}). The partition function $f_l(x)$ with $l=3$ is used in contrast to that with $l=1$ in the original version (see \citealt{Brasser2015}).

The benchmarking is conducted on one of the general purpose compute nodes with dual AMD EPYC 7742 at 2.25 GHz in the HPC2021 cluster running CentOS 8.5 at the University of Hong Kong (HKU). Executables are compiled with GFortran 11.2 at the default optimization level.

\begin{figure}[t]
	\centering
	\includegraphics{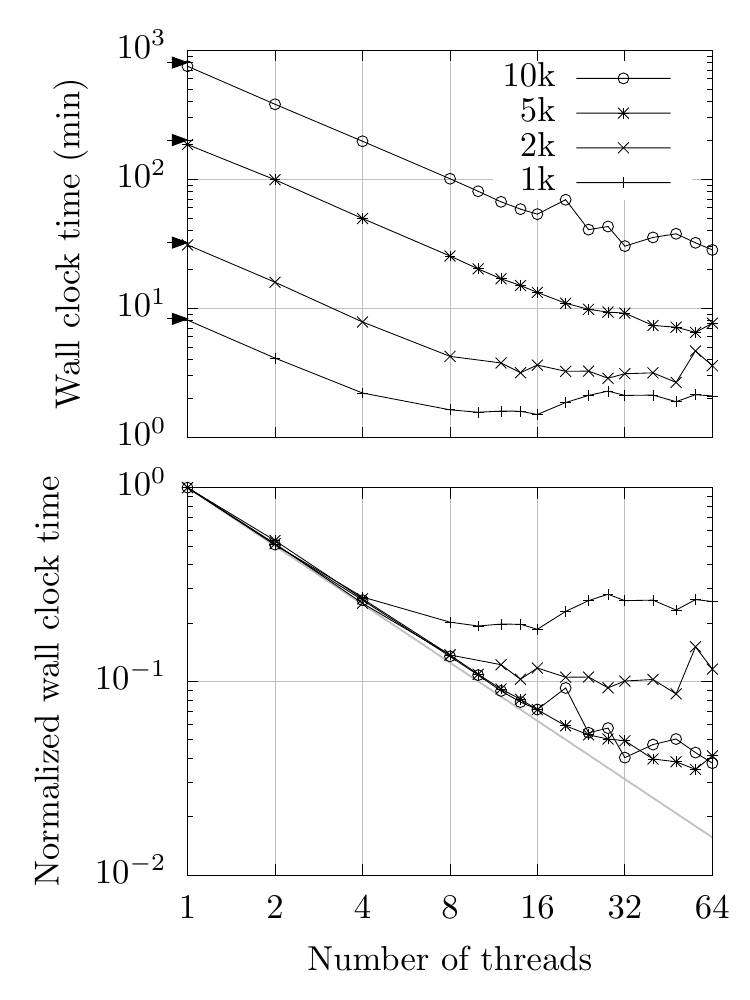}
	\caption{Benchmarking results of SyMBAp with the 4 conditions. Top: Wall clock times with respect to the number of threads. Arrows on the $y$-axis indicate the wall clock times using the original SyMBA. Bottom: Normalized wall clock times for SyMBAp.}
	\label{bm}
\end{figure}

Figure \ref{bm} summarizes the benchmarking results. With one thread, SyMBAp shows a slight performance gain compared to SyMBA, which is due to other optimizations implemented in the course of this work. The largest fractional reduction in wall clock time is achieved for condition ``5k'', which shows a speedup of 30.8 times with 56 threads. Generally, the scalability is better for larger $N$, because the overhead from fork-join parallelism becomes less significant relative to the parallelized operations. Thanks to the availability of high-core-count nodes, optimal performance has been achieved for the setups benchmarked, where additional cores cannot further boost the performance significantly.

To confirm the accuracy of the implementation, another planetesimal disk is generated with $100$ planetesimals of $10^{-2} M_\earth$ with separation of 5 Hill radii. The innermost particle is placed at 5 AU. The eccentricities and inclinations are drawn in the manner described above. The system is integrated with full interactions using SyMBA and SyMBAp for $10^4$ years with time step size of $0.1$ year. Throughout the integrations, no merger occurred and no particle is removed. Differences in the energy and angular momentum errors of SyMBA and SyMBAp are consistent with changes in the sequence of operations and rounding error accumulation. The test shows the preservation of the symplectic nature and the accurate implementation of SyMBAp.

\section{Summary} \label{sec:summary}
In this work, we have presented a parallelized $N$-body code named SyMBAp. We have performed extensive benchmarking and tests on various computers to confirm its portability, scalability and accuracy, and some representative examples are shown here. The source code of SyMBAp is available at \url{https://github.com/tommylauch/SWIFT_SyMBAp_pub} and archived at \url{https://doi.org/10.5281/zenodo.7819840}.

\begin{acknowledgments}
This work was supported by Hong Kong RGC grant 17306720, a postgraduate studentship and a seed fund for basic research at HKU.
Parts of the computations were performed using research computing facilities offered by HKU Information Technology Services.
\end{acknowledgments}

\bibliography{symba5p}
\bibliographystyle{aasjournal}

\end{document}